\documentclass[twocolumn,amsmath,amssymb,floatfix,superscriptaddress]{revtex4-1}
\usepackage{bm,graphicx,url,epsf,color}
\usepackage[french]{babel}
\usepackage{wasysym}
\usepackage{MnSymbol}

\begin{document}

\title{Macroscopic electron quantum coherence in a solid-state circuit}

\author{H. Duprez}
\email{These authors contributed equally to this work.}
\affiliation{Centre de Nanosciences et de Nanotechnologies (C2N), CNRS, Univ Paris Sud, Universit\'e Paris-Saclay, 91120 Palaiseau, France}
\author{E. Sivre}
\email{These authors contributed equally to this work.}
\affiliation{Centre de Nanosciences et de Nanotechnologies (C2N), CNRS, Univ Paris Sud, Universit\'e Paris-Saclay, 91120 Palaiseau, France}
\author{A. Anthore}
\affiliation{Centre de Nanosciences et de Nanotechnologies (C2N), CNRS, Univ Paris Sud, Universit\'e Paris-Saclay, 91120 Palaiseau, France}
\affiliation{Univ Paris Diderot, Sorbonne Paris Cit\'e, 75013 Paris, France}
\author{A. Aassime}
\affiliation{Centre de Nanosciences et de Nanotechnologies (C2N), CNRS, Univ Paris Sud, Universit\'e Paris-Saclay, 91120 Palaiseau, France}
\author{A. Cavanna}
\affiliation{Centre de Nanosciences et de Nanotechnologies (C2N), CNRS, Univ Paris Sud, Universit\'e Paris-Saclay, 91120 Palaiseau, France}
\author{A. Ouerghi}
\affiliation{Centre de Nanosciences et de Nanotechnologies (C2N), CNRS, Univ Paris Sud, Universit\'e Paris-Saclay, 91120 Palaiseau, France}
\author{U. Gennser}
\affiliation{Centre de Nanosciences et de Nanotechnologies (C2N), CNRS, Univ Paris Sud, Universit\'e Paris-Saclay, 91120 Palaiseau, France}
\author{F. Pierre}
\email[e-mail: ]{frederic.pierre@c2n.upsaclay.fr}
\affiliation{Centre de Nanosciences et de Nanotechnologies (C2N), CNRS, Univ Paris Sud, Universit\'e Paris-Saclay, 91120 Palaiseau, France}


\begin{abstract}
The quantum coherence of electronic quasiparticles underpins many of the emerging transport properties of conductors at small scales \cite{Nazarov2009}.
Novel electronic implementations of quantum optics devices are now available \cite{vanWees1989b,Ji2003,Neder2007,Yamamoto2012,Bocquillon2013a,Dubois2013} with  perspectives such as `flying' qubit manipulations \cite{Bertoni2000,Ionicioiu2001,Stace2004,Glattli2017,Bauerle2018}.
However, electronic quantum interferences in conductors remained up to now limited to propagation paths shorter than $30\,\mu$m, independently of the material \cite{Pierre2003,Niimi2010,Miao2007}.
Here we demonstrate strong electronic quantum interferences after a propagation along two 0.1\,mm long  pathways in a circuit.
Interferences of visibility as high as $80\%$ and $40\%$ are observed on electronic analogues of the Mach-Zehnder interferometer of, respectively, $24\,\mu$m and $0.1\,$mm arm length, consistently corresponding to a $0.25\,$mm electronic phase coherence length.
While such devices perform best in the integer quantum Hall regime at filling factor 2 \cite{Roulleau2008a,Litvin2008,Gurman2016}, the electronic interferences are restricted by the Coulomb interaction between copropagating edge channels \cite{Levkivskyi2008,Roulleau2008b}.
We overcome this limitation by closing the inner channel in micron-scale loops of frozen internal degrees of freedom \cite{Altimiras2010,Cabart2018}, combined with a loop-closing strategy providing an essential isolation from the environment.
\end{abstract}

\maketitle

\section{Introduction}

Ballistic electrons allow for advanced quantum manipulations at the single electron level in circuits, in the spirit of the manipulation of photons in quantum optics \cite{Bocquillon2014,Glattli2017,Bauerle2018}.
Perspectives notably include a different paradigm for quantum information processing, with a non-local architecture based on `flying' qubits encoded for example by the presence or absence of an electron within a propagating wave packet \cite{Bertoni2000,Ionicioiu2001,Stace2004,Bautze2014,Glattli2017,Bauerle2018}.
Electronic edge states topologically protected against disorder constitute promising solid-state platforms.
In particular, the emblematic chiral edge channels propagating along a two-dimensional (2D) conductor in the quantum Hall regime are generally considered ideal 1D conductors.  
Their analogy with light beams, their in-situ tunability by field effect and the availability of single-electron emitters were exploited to implement the electronic analogues of optical devices, such as the interferometers of types Fabry-Perot \cite{vanWees1989b}, Mach-Zehnder \cite{Ji2003}, Hanbury-Brown and Twiss \cite{Neder2007} and Hong-Ou-Mandel \cite{Bocquillon2013a}.
In contrast to photons, the Coulomb interaction between charged electrons provides a natural correlation mechanism to realize e.g. CNOT gates \cite{Bertoni2000,Ionicioiu2001,Glattli2017,Bauerle2018}.
However, the same Coulomb interaction generally entangles the propagating electrons efficiently with numerous degrees of freedom, including the surrounding electrons, which gives rise to quantum decoherence \cite{Nazarov2009} (see \cite{Duprez2019b} for a notable exception).

In practice, the maximum electron phase coherence length $L_\phi$ was previously found to reach remarkably similar values at the lowest accessible temperatures  in very diverse systems, from diffusive metal ($L_\phi\simeq20\,\mu$m reported in \cite{Pierre2003} at 40\,mK) to near ballistic two-dimensional electron gas ($L_\phi\simeq20\,\mu$m reported in \cite{Niimi2010} at 30\,mK) and graphene ($L_\phi\simeq3-5\,\mu$m estimated in \cite{Miao2007} at 260\,mK).
Along the ballistic quantum Hall edge channels of specific interest for electron quantum optics, $L_\phi\simeq24\,\mu$m was demonstrated at 20\,mK \cite{Roulleau2008a} at the most advantageous magnetic field tuning corresponding to filling factor $\nu=2$ in a Ga(Al)As 2D electron gas.
We also point out two promising findings: an important temperature robustness of small conductance oscillations measured across a 6\,$\mu$m long Ga(Al)As device, from which a large value of $L_\phi\sim86\,\mu$m was indirectly inferred \cite{Yamamoto2012}; and conductance oscillations of very high visibility along a graphene pn junction \cite{Wei2017}.
Here, we establish a macroscopic electron phase coherence length, of $0.25$\,mm, achieved along quantum Hall channels by nano-circuit engineering.

At low temperatures, short-range electron-electron interactions within the {\it same} chiral edge channel of the integer quantum Hall regime are predicted to increase the electrons' propagation velocity, but not to limit their coherence \cite{Giamarchi2003,Levkivskyi2008}.
The dominant dephasing mechanism is generally attributed to the interaction between electrons located in {\it adjacent} edge channels \cite{Levkivskyi2008,Roulleau2008b} (except at $\nu=1$ and fractional filling factors where the stronger decoherence \cite{Litvin2008,Gurman2016} is not clearly understood).
This picture is established by complementary signatures including energy transfers \cite{leSueur2010,Itoh2018}, charge fractionalization \cite{Bocquillon2013b,Inoue2014b,Hashisaka2017} and Hong-Ou-Mandel characterizations \cite{Marguerite2016}.
However, additional dissipative mechanisms yet unidentified were also evidenced experimentally, even in the most canonical $\nu=2$ case \cite{leSueur2010,Bocquillon2013b,Itoh2018}.
In this work, we demonstrate a circuit design strategy that very efficiently suppresses the essential decoherence mechanisms.
 
\section{Nanoengineering the phase coherence length}
The electronic version of the Mach-Zehnder interferometer (MZI, schematically depicted in Fig.~1(a)) essentially consists in a quantum Hall edge channel following two separate paths, and in two quantum point contacts (QPC) used as tunable beam splitters \cite{Ji2003}.
The quantum Hall regime is realized in a Ga(Al)As 2D electron gas immersed in a perpendicular magnetic field of 4.3\,T corresponding to a filling factor $\nu=2$, with two copropagating edge channels.
The interfering MZI paths involve only the outer edge channel (thick black lines in Fig.~1(a)).
The two beam splitter QPCs are formed by field effect using split gates (colored orange in Fig.~1(a); with suspended bridges to contact the top parts). 
The quantum phase difference between the two paths is proportional to the enclosed magnetic flux.
It is here controlled by fine-tuning the lower edge path with the voltage $V_\mathrm{pl}$ applied to a lateral plunger gate (colored green in Fig.~1(a),(b)).
The quantum interferences are evidenced by sweeping $V_\mathrm{pl}$, from the resulting oscillations of the transmitted current impinging on the metallic electrode labeled D in Fig.~1(a).
Their energy dependence, with respect to the bias voltage $V_\mathrm{dc}$ applied to the source electrode, is obtained from a concomitant noise in the transmitted current.
The second MZI output is connected to the central metallic electrode (elongated yellow disk in Fig.~1), which is electrically grounded through a suspended bridge.
In contrast to previous MZI implementations, our devices include two long surface gates (light gray in Fig.~1(a),(b)) with a particular comb shape with both shafts and teeth placed over the 2D electron gas.
This is essential for the presently demonstrated strong increase of the electron coherence.
As illustrated in Fig.~1(a), these gates can be biased to form inner channel loops along the interfering outer edge channel paths.
In order to unambiguously demonstrate and accurately measure very large phase coherence lengths, we fabricated two MZI with extraordinarily long symmetric arms of length $L\simeq24\,\mu$m (Fig.~1(b)) and $0.1\,$mm (Fig.~1(c)).
For a straightforward comparison at different $L$, the two devices were made concurrently (a few millimeters away on the same chip), with identical designs except for the length of the elongated central area, and were simultaneously cooled-down to 10\,mK.

\begin{figure}
\centering\includegraphics[width=86mm]{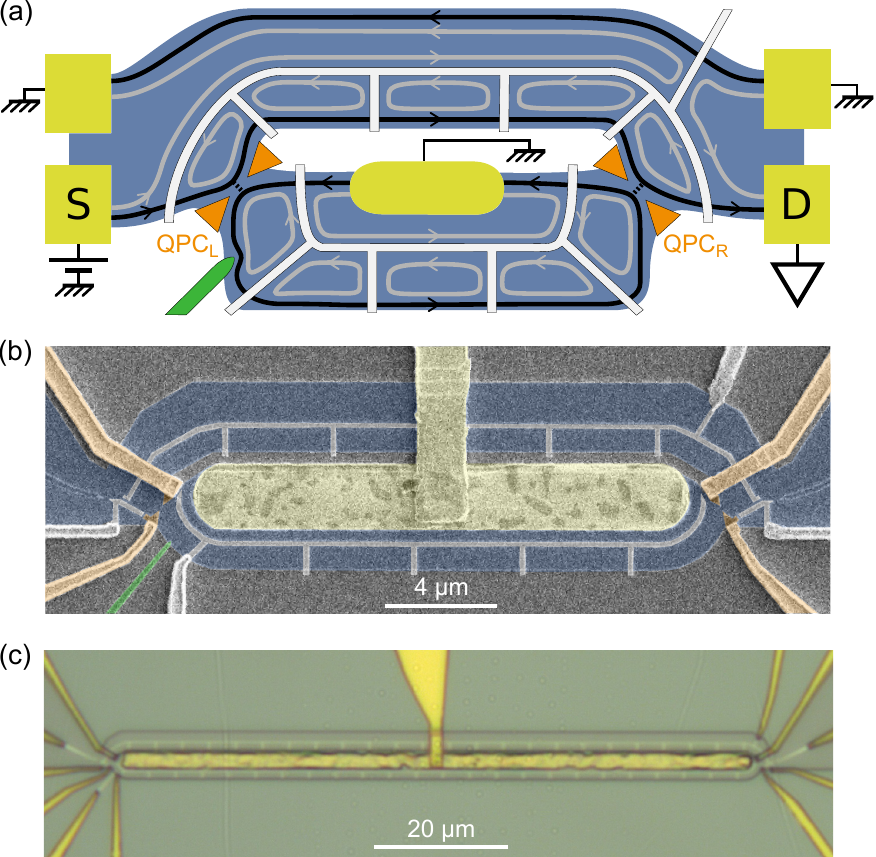}
\caption{
Nano-circuit engineering of electronic coherence.
(a) Sample schematic.
Two chiral edge channels (black and gray lines with arrows) propagate along a 2D electron gas (blue) set in the integer quantum Hall regime at filling factor $\nu=2$. 
The outer channel (black) follows two separate paths between tunable beam splitters implemented by quantum point contacts (orange), thereby forming a Mach-Zehnder interferometer.
The inner edge channel (gray) can be closed into well-separated loops, with specific comb-shaped gates (light gray) voltage biased to reflect only this channel.
Sweeping the voltage on a lateral plunger gate (green) results in MZI oscillations of the current transmitted from source (S) to detector (D).  
(b) Colored scanning electron micrograph of the sample with MZI arms of symmetric length $L\simeq24\,\mu$m.
(c) Optical image of the $L\simeq0.1$\,mm MZI.
The inner edge channel loops have nominally identical perimeters of $9\,\mu\mathrm{m}$, except one of $5\,\mu\mathrm{m}$ for the lower left loop of each sample.
}
\label{fig-sample}
\end{figure}

\begin{figure*}
\centering\includegraphics[width=178mm]{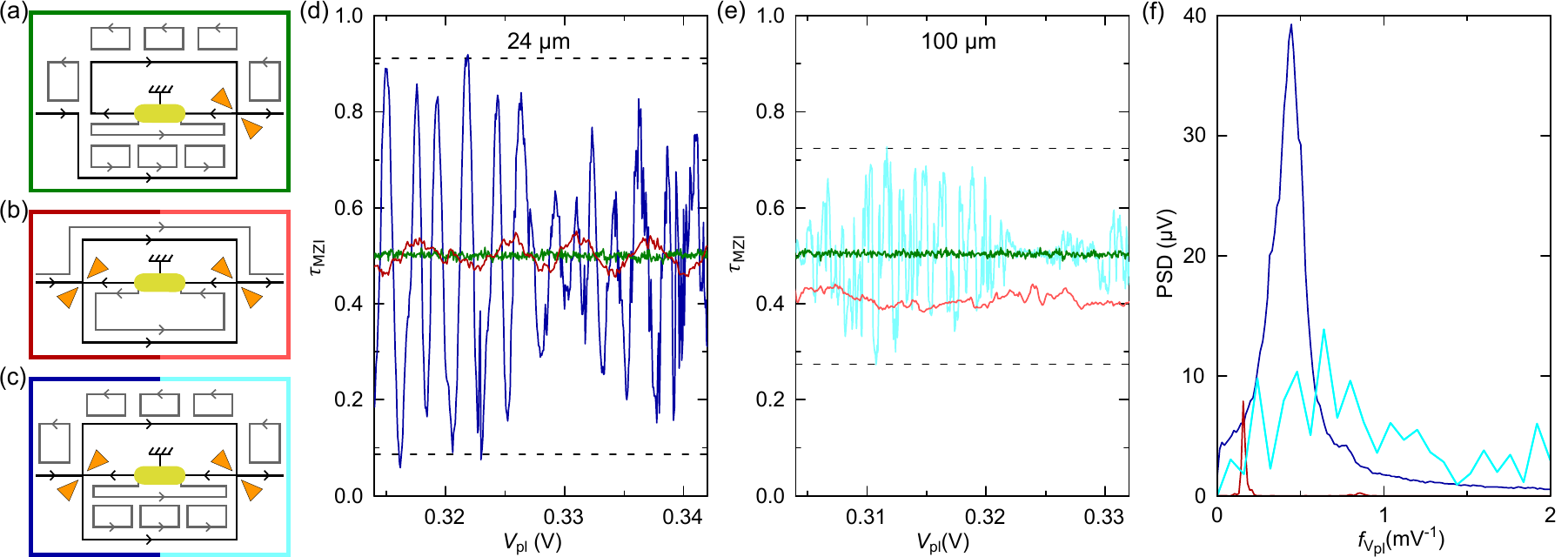}
\caption{
\textbf{Quantum oscillations}. 
(a),(b),(c) Schematics of the different configurations.
(d),(e) Continuous lines show, versus plunger gate voltage $V_\mathrm{pl}$, the measured fraction $\tau_\mathrm{MZI}$ of current transmitted from S to D along the outer channel of the $L\simeq24\,\mu$m (d) and $0.1\,$mm (e) MZI (same color as the box enclosing the corresponding schematic in panel (a), (b) or (c); darker shade for the shorter device).  
Horizontal dashed lines display the predicted $\tau_\mathrm{MZI}$ extrema for the same $L_\phi=0.25\,$mm in both MZI.
(f) Continuous lines show the power spectral density of $\tau_\mathrm{MZI}(V_\mathrm{pl})$, determined along large $V_\mathrm{pl}$ sweeps (extending between 50 and 80\,mV) measured several times (same color code as in panels (d),(e)).
For the challenging case of $L\simeq0.1\,$mm in configuration (c) (light blue line), the Fourier analysis was restricted to plunger gate voltage windows exhibiting oscillations larger than $66\%$ of their maximum amplitude.
}
\label{fig-interference}
\end{figure*}

How can $L_\phi$ be increased?
It was initially shown that most of the electrons' energy relaxation can be frozen within the outer edge channel at $\nu=2$ (along a 8\,$\mu$m path), by closing into a loop the inner channel \cite{Altimiras2010}.
This was explained by the electronic levels' quantization within the loop, which effectively quenches the phase space for inelastic collisions with the inner loop's electrons (for a level spacing larger than the available energy) \cite{Altimiras2010,Cabart2018}.
As inelastic collisions also result in decoherence, a similar approach was subsequently tested on $L_\phi$ using an electronic MZI \cite{Huynh2012}.
However, the increase in $L_\phi$ by forming inner channel loops was limited to a factor of two \cite{Huynh2012}, relatively modestly compared to the freezing of energy relaxation.
Our conjecture is that the weaker impact on $L_\phi$ reflects a fundamental design limitation in the MZI implementation of \cite{Huynh2012}, where an otherwise negligible coupling between two different \textit{outer} edge channels could be mediated by the rigid displacements of the inner loops. This provides an additional mechanism for both decoherence and energy relaxation: even if the inner loops' electronic degrees of freedom are not excited, the loops' presence can strongly enhance the capacitive coupling between different propagative edge channels adjacent to separate portions of the same loops.
The present MZI implementation suppresses this mechanism while preserving a 2D bulk at $\nu=2$, through a gate design allowing for a much larger separation of the inner loops from additional quantum Hall channels (see Fig.~4 for an illustration, and Appendix section 2 for further discussion).

\section{Quantum oscillations versus loop formation}

We present in Fig.~2 illustrative MZI oscillations, versus plunger gate voltage $V_\mathrm{pl}$ (a positive bias of $+0.35$\,V was applied during cooldown).
The displayed $\tau_\mathrm{MZI}$ corresponds to the transmission probability across the MZI, from source S to detector D.
It is given by the fraction measured at the electrode D of the current injected into the outer edge channel at the electrode S.
The two $L\simeq24\,\mu$m and $0.1\,$mm MZI are each tuned in three different configurations (Fig.~2(a),(b),(c)).
The green lines in Fig.~2(d),(e) are data obtained with both devices set in the configuration shown in Fig.~2(a). 
Their flatness demonstrates directly, in the presence of inner channel loops, the absence of $\tau_\mathrm{MZI}$ oscillations when all the transmitted current goes through a single MZI arm (the lower arm; in this specific case $\tau_\mathrm{MZI}=\tau_\mathrm{QPC}^\mathrm{R}$ since $\tau_\mathrm{QPC}^\mathrm{L}=1$).
The red and blue lines in Fig.~2(d),(e) are obtained with both QPC beam splitters set to half transmission probability for the outer edge channel ($\tau_\mathrm{QPC}^\mathrm{L}\simeq\tau_\mathrm{QPC}^\mathrm{R}\simeq0.5$, the inner edge channel being always fully reflected at the QPCs) in the configurations illustrated in Fig.~2(b),(c).
In the conventional MZI configuration (no loops, Fig.~2(b)), small oscillations of period $6.4$\,mV are observed only on the $L\simeq24\,\mu$m device (dark red lines in Fig.~2(d),(f)). 
Their visibility $\mathcal{V}\equiv(\tau_\mathrm{MZI}^\mathrm{max}-\tau_\mathrm{MZI}^\mathrm{min})/(\tau_\mathrm{MZI}^\mathrm{max}+\tau_\mathrm{MZI}^\mathrm{min})\approx6\%$ corresponds to a typical phase coherence length value of $L_\phi\simeq17\,\mu$m (despite a relatively low temperature $T\simeq10\,$mK) obtained from the standard relationship for a symmetric MZI:
\begin{equation}
\mathcal{V}=4\sqrt{\tau_\mathrm{QPC}^\mathrm{R}(1-\tau_\mathrm{QPC}^\mathrm{R})\tau_\mathrm{QPC}^\mathrm{L}(1-\tau_\mathrm{QPC}^\mathrm{L})}\exp\left(\frac{-2L}{L_\phi}\right), \label{eqLphi}
\end{equation}
which assumes a perfect absorption of the outer edge channel by the central metallic contact connected to electrical ground (separately checked, Appendix).
In contrast, for the $L\simeq0.1\,$mm device, no oscillations can be detected without inner channel loops as expected from Eq.~\ref{eqLphi} ($\mathcal{V}\approx10^{-5}$ calculated with $L=0.1\,$mm and $L_\phi=17\,\mu$m).
Instead, we observe a slowly evolving $\tau_\mathrm{MZI}$, which is markedly below $0.5$.
This low mean value reflects the tunneling of electrons from outer to inner edge channels, which becomes significant over such a long propagation distance.
As a result, a larger (smaller) fraction of the current injected into the outer edge channel is absorbed by the grounded central ohmic contact (detected at D).
Specific measurements of the tunneling between copropagating channels are discussed in the Appendix (section 5).

With inner channel loops formed (Fig.~2(c)), high amplitude oscillations of maximum visibility $\mathcal{V}\approx80\%$ and $40\%$ are observed for the $L\simeq24\,\mu$m and $0.1\,$mm MZI, respectively.
Their sinusoidal shape is however perturbed by jumps as well as amplitude modulations, which are attributed to fluctuators such as moving charges in the MZI vicinity.
A sudden variation in surrounding charges would indeed appear as a phase jump.
In contrast, relatively rapid fluctuations with respect to the experimental integration time ($\sim1$\,s), but slow with respect to the electron quantum coherence time, would artificially reduce the amplitude of MZI oscillations, below their intrinsic value limited by $L_\phi$ according to Eq.~\ref{eqLphi}.
As illustrated with the emblematic single electron transistor, individual charge fluctuators are usually influenced by surrounding gate voltages.
Accordingly, we observe modulations of the phase jump density and of the amplitude of oscillations with gate voltages.
Note that two sources of moving charges are specific to the present MZI implementation with inner channel loops: ({\it{i}}) the voltage bias applied to the very long surface gates used to form the loops, and ({\it{ii}}) jumps in the number of electrons within each of the many inner channel loops (from the possible tunneling of electrons between outer channel and inner loops).
We now further establish, by a train of evidence, that the large oscillations observed with inner channel loops result from the quantum interferences between the two MZI paths, and that their maximum visibility accurately reflects $L_\phi$.

\section{Oscillation characterization}

\begin{figure*}
\centering\includegraphics[width=129mm]{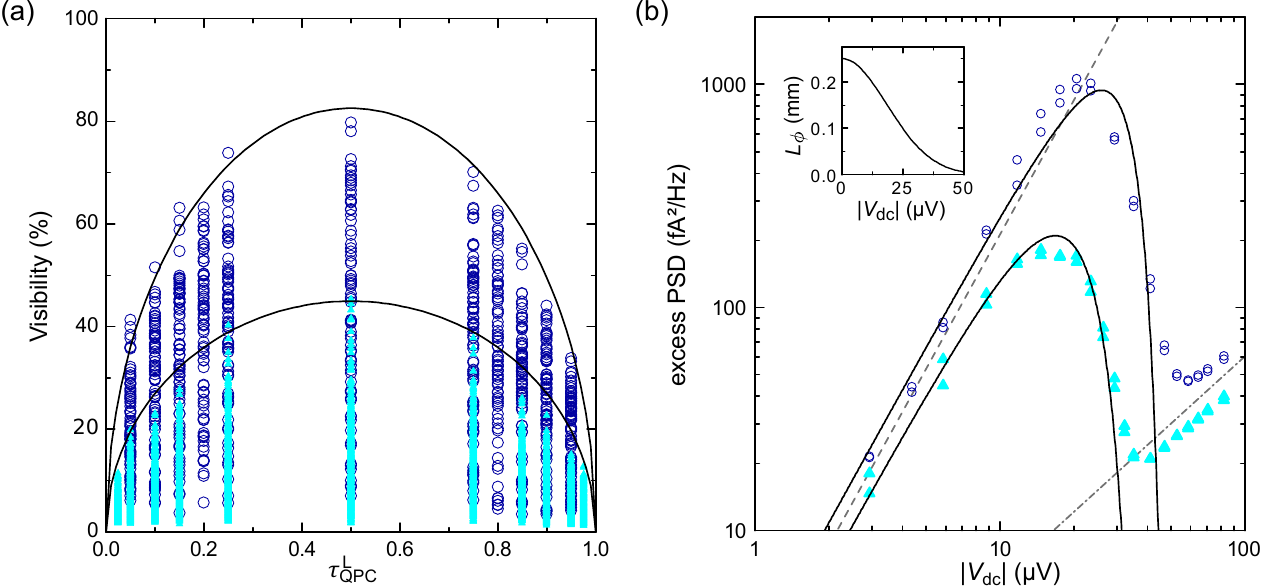}
\caption{
Beam splitter and bias voltage tunings. 
Open (full) symbols are data points obtained on the $L\simeq24~(100)\,\mu$m MZI.
(a) The local quantum oscillations' visibility in the presence of inner channel loops (Fig.~2(c)), separately extracted period per period along large $V_\mathrm{pl}$ sweeps, is displayed as symbols versus the transmission probability $\tau_\mathrm{QPC}^\mathrm{L}$ of the outer channel across the left QPC (at fixed $\tau_\mathrm{QPC}^\mathrm{R}\simeq0.5$).
Continuous lines are Eq.~\ref{eqLphi}'s predictions for $L_\phi=0.25\,$mm with $L=24\,\mu$m or $0.1\,$mm.
(b) The excess power spectral density of temporal fluctuations in the transmitted MZI current, with respect to zero dc bias and averaged in $V_\mathrm{pl}$, is shown versus source (S) dc voltage $V_\mathrm{dc}$.
The gray straight lines represent a quadratic (dashed) and linear (dash-dotted) increase.
The black continuous lines in the main panel display the noise contribution from phase fluctuations calculated with $L_\phi(V_\mathrm{dc})=(0.25\,\mathrm{mm})\times\exp\left[-(V_\mathrm{dc}/26\,\mu\mathrm{V})^2\right]$ (shown in the inset).
}
\label{fig-nonEq}
\end{figure*}

First, a well-defined plunger gate voltage period of $2.2$\,mV is observed for the smaller $L\simeq24\,\mu$m MZI, as directly evidenced from the power spectral density (dark blue lines in Fig.~2(d),(f)). 
A compatible but broader oscillation periodicity can also be perceived for the $L\simeq0.1\,$mm MZI, but only if the FFT analysis is restricted to plunger gate voltage windows where the oscillation amplitude is relatively large (light blue line in Fig.~2(f)).
The period for $L\simeq24\,\mu$m with loops is shorter than without, as expected from the stronger influence of the plunger gate voltage.
This is a consequence of the quenched screening from isolated inner channel loops, hosting a discrete number of electrons, as compared to a copropagative inner channel.
It also implies that any nearby moving charges will have a stronger impact on the MZI quantum phase.

Second, as shown in Fig.~3(a), the maximum oscillation visibility (highest symbols) follows the hallmark MZI signature $\sqrt{\tau_\mathrm{QPC}^\mathrm{L}(1-\tau_\mathrm{QPC}^\mathrm{L})}$ (continuous lines), when varying the outer edge channel transmission probability across the left QPC beam splitter $\tau_\mathrm{QPC}^\mathrm{L}$.
For this purpose, we have measured $\tau_\mathrm{MZI}(V_\mathrm{pl})$ over many periods on both devices, and for various settings of $\tau_\mathrm{QPC}^\mathrm{L}$ at fixed $\tau_\mathrm{QPC}^\mathrm{R}\simeq0.5$ (Appendix).
Each symbol in Fig.~3(a) (full and open corresponding to the $L\simeq24\,\mu$m and $0.1\,$mm MZI, respectively) displays the `locally' extracted visibility of the oscillations, obtained by analyzing a restricted plunger gate voltage window of one period ($2.2$\,mV). 
The close agreement between highest data points and MZI expectations confirms that the observed oscillations result from the two-path quantum interferences.

Third, we find a quantitative data/theory agreement with the {\it{same}} $L_\phi\approx0.25$\,mm for both devices, despite a factor of four in their size.
The continuous lines in Fig.~3(a) are calculated using Eq.~\ref{eqLphi} with $L_\phi=0.25$\,mm, the corresponding MZI length $L=24\,\mu$m or $0.1\,$mm, and $\tau_\mathrm{QPC}^\mathrm{R}=0.5$.
This provides a strong evidence that the measured maximum `local' visibility closely captures the intrinsic MZI visibility, determined solely by $L_\phi$ (note that $L_\phi$ would otherwise be underestimated).

Fourth, as shown in Fig.~3(b), out-of-equilibrium measurements of the transmitted current noise around $0.86\,$MHz further confirm the presence of MZI interferences accompanied by phase fluctuations, and allow probing the energy dependence of $L_\phi$.
The displayed data points represent measurements of the excess power spectral density of the current impinging on the electrode D, versus the dc bias voltage $V_\mathrm{dc}$ applied to the source electrode S.
MZI phase variations, such as those produced by nearby charge fluctuators, are expected to manifest as a quadratic increase of the noise power at low $V_\mathrm{dc}$ (see Appendix and \cite{Marquardt2004}), as experimentally observed.
At larger bias, the generally expected reduction of $L_\phi$ also progressively diminishes the influence of the quantum phase and, consequently, the current noise induced by phase fluctuations.
Experimentally, such a collapse is observed and can be accounted for using the {\it{same}} $L_\phi(V_\mathrm{dc})$ for both devices:
the two black continuous lines (main panel) are calculations based on Eq.~\ref{eqLphi} (Appendix, see Eq.~\ref{eqPSDnoshotnoise}) using the empirically determined $L_\phi=(0.25\,\mathrm{mm})\times\exp\left[-(V_\mathrm{dc}/26\,\mu\mathrm{V})^2\right]$ (shown in inset).
Ultimately, a linear noise increase is recovered as expected for the shot noise contribution \cite{Marquardt2004} (Appendix).

\section{Discussion}

The large phase coherence length presently achieved provides information for the design of novel quantum Hall devices.
It sets an upper bound to possibly relevant decoherence mechanisms along the quantum Hall edges, besides the dominant inter-channel coupling, and narrows down the mechanisms for a frequently observed but still mysterious additional dissipation \cite{leSueur2010,Bocquillon2013b,Itoh2018,Krahenman2019}.

We establish that nearby metallic gates are completely compatible with large phase coherence lengths, despite the presence of many diffusive electrons.
Note their beneficial screening of the long-range part of Coulomb interaction (to approximately $\sim3.5\,\mu$m, the loop-gates' period, whether the loops are formed or not), which could otherwise provide an effective decoherence mechanism \cite{Sukhorukov2007,Chalker2007,Kovrizhin2010,Schneider2011} as well as an unwanted coupling to spurious low energy modes and distant channels \cite{Hashisaka2012,Tu2018,Krahenman2019}.
In practice, a strong capacitive shortcut (100\,nF) was included at the low temperature end of the electrical lines controlling the gates of our samples, in order to further suppress both extrinsic and thermal noise sources.

We also find that the additional neutral modes predicted for a realistic smooth confinement potential at the edge \cite{MacDonald1993,Chamon1994,Aleiner1994} can essentially be ignored.
Either these neutral modes are missing in the outer channel along our etched-defined edges, or they are very weakly coupled to the usual charge mode of the same channel.
This is consistent with thermal conductance measurements across narrow constrictions perfectly transmitting one or several quantum Hall channels at integer bulk filling factors, where the extra heat transfer that would be expected from additional edge modes was not observed \cite{Jezouin2013b,Banerjee2017,Sivre2018}.

Finally, we mention that the two-dimensional quantum Hall bulk does not provide here a substantial path to quantum decoherence, at least when broken into small areas of a few micron squares (within the inner channel loops) and with the long range part of Coulomb interaction screened by metallic gates.
This contrasts with the observations of an unexpected heat flow away from the edge at lower filling factors \cite{Altimiras2012,Venkatachalam2012,Inoue2014a} and of a long-distance capacitive coupling across the two-dimensional bulk \cite{Hashisaka2012,Tu2018}.

\section{Conclusion}

We have demonstrated that the electron quantum coherence in solid state circuits can be extended to the macroscopic scale, by strongly suppressing through circuit nano-engineering the dominant decoherence mechanism.
The present implementation on quantum Hall edge channels is particularly well suited for the coherent control and long distance entanglement of propagative electrons.
Future optimizations include the understanding and suppression of the slow electron phase fluctuations here often, although not systematically, observed.
Our work gives access to electron quantum optics devices of a higher complexity level, in line with the direction taken by this field of research \cite{Yamamoto2012,Bocquillon2014,Glattli2017,Bauerle2018}. 
More generally, increasing the electron phase coherence is essential to progress toward functional quantum devices involving multiple quantum manipulations, such as information processing with electronic flying qubits.

\begin{acknowledgements}
This work was supported by the French RENATECH network, the national French program `Investissements d'Avenir' (Labex NanoSaclay, ANR-10-LABX-0035) and the French National Research Agency (project QuTherm, ANR-16-CE30-0010-01).

E.S. and H.D. performed the experiment and analyzed the data with inputs from A.Aa., A.An. and F.P.;
F.P. fabricated the sample with inputs from E.S and H.D.;
A.C., A.O. and U.G. grew the 2DEG;
F.P. led the project and wrote the manuscript with inputs from A.Aa., A.An., E.S., H.D. and U.G.
\end{acknowledgements}

\section*{Appendix} 

\subsection*{1. Samples}
Both samples are made of the same Ga(Al)As heterojunction hosting a two-dimensional electron gas of mobility $10^6\,\mathrm{cm}^2\mathrm{V}^{-1}\mathrm{s}^{-1}$ and density $2.5\,10^{11}\,\mathrm{cm}^{-2}$, located $105$~nm underneath the surface. 
They were nano-fabricated by e-beam lithography, dry etching and metallic deposition. 
The central metallic electrode (nickel [30\,nm], gold [120\,nm] and germanium [60\,nm]) forms an ohmic contact with the 2DEG, obtained by thermal annealing (at 440\,$^\circ$C for 50\,s), and is set to electrical ground through a suspended bridge.
The two arms of each MZI were designed to be as symmetric as possible, such that the thermal smearing of the visibility induced by an asymmetry would remain negligible by a large margin as previously observed \cite{Roulleau2008a,Huynh2012}.
The elongated shape of the central area was chosen to limit the overall magnetic flux enclosed between the two arms, and hence the effect of environmental magnetic noise (e.g. from the pulse tube vibrations) on the particularly sensitive MZI phase in these very large devices.
Note that a positive bias voltage of $+0.35$\,V was applied to all used gates during cooldown.
This is a widespread procedure in Ga(Al)As devices to reduce the charge noise induced by biasing the gates, although it is probably not essential here due to the relatively low bias voltages used to form inner channel loops.\\

\begin{figure}
\centering\includegraphics [width=86mm]{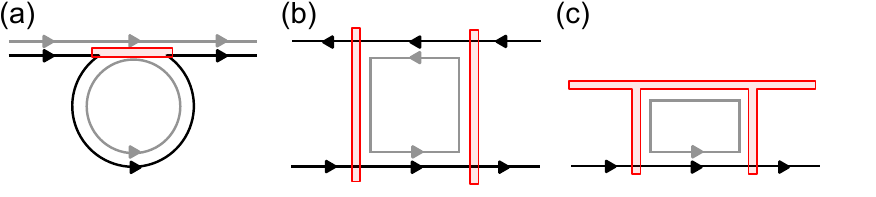}
\caption{
Loop design.
Inner loop design in previous energy relaxation experiment \cite{Altimiras2010} (a), previous MZI experiment \cite{Huynh2012} (b) and in the present implementation (c).
The outer (inner) edge channel is represented by a black (gray) line.
A schematic of the gates used to reflect the inner edge channel is displayed in red. 
}
\end{figure}

\subsection*{2. Loop gate design} 
Figure~4 recapitulates the different kinds of inner channel loops in the energy relaxation experiment \cite{Altimiras2010} (one inner loop enclosed only by the outer channel, see panel (a)), in the first MZI implementation \cite{Huynh2012} (inner loops enclosed by a metallic gate, the MZI outer channel and an other counter-propagating outer channel, see panel (b)) and in the present MZI implementation (inner loops enclosed by a metallic gate and the MZI outer channel, see panel (c)).
Now focusing on the present implementation, the gates' width of 200\,nm reflects a compromise between the separation with additional quantum Hall channels on the other side of the gates, which should be sufficiently large to result in a negligible coupling, and the wish to limit the $\nu=1$ area underneath the gates, as very weak interferences are often observed if the whole 2D bulk is set to $\nu=1$ (either by tuning $B$ without gates or using a broad top gate fully covering the 2D bulk, see e.g. \cite{Gurman2016}).
The distance between inner channel loops and propagative (inner) quantum Hall channel on the other side of the gates (opposite to the MZI outer channel) should therefore be larger than 200\,nm.
This is more than one order of magnitude larger than the narrow incompressible strip normally separating adjacent edge channels (typically 10\,nm \cite{Chklovskii1992}).
The loops' perimeter should also be chosen small enough such that the separation between the quantized electronic levels is larger than the available energy $\sim k_\mathrm{B}T$.
Assuming a typical drift velocity between $10^4$ and $10^5$\,m/s along the sample edges, we find that the 9\,$\mu$m loop perimeter corresponds to a level spacing within 4.6 and 46\,$\mu e$V, always larger than the thermal energy ($3k_\mathrm{B}T\simeq2.6\,\mu e$V at 10\,mK) and comparable to the characteristic $26\,\mu$V dc bias voltage over which $L_\phi(V_\mathrm{dc})$ is found to decrease (Fig.~3(b)).
Finally, the gates were designed elongated to minimize their overlap with the outer MZI edge channel, as at these locations their capacitive coupling is maximal and the lateral edge confinement is modified. 
Note also that one should be particularly careful about the electrical noise introduced by the measurement lines connected to the very long gates used to form the inner channel loops.
These are indeed much more strongly coupled to the MZI phase than typical lateral plunger gates, due to their very long size and because the inner loop efficiently mediate the capacitive coupling between metallic gate and MZI outer edge channel.

\subsection*{3. Experimental setup} 
The two, simultaneously cooled devices are thermally anchored to the mixing chamber of a cryofree dilution refrigerator.
Electrical lines connected to the samples include multiple filters and thermalization stages.
Note the important RC filter (200\,k$\Omega$, 100\,nF) implemented at base temperature on the lines connected to the gates, including the long gates used to form the inner channel loops.
Spurious high-frequency radiations are screened by two shields at base temperature.
The fraction of transmitted current $\tau_\mathrm{MZI}$ is measured with lock-ins, at a frequency below 200\,Hz and using an effective integration time close to one second per point (corresponding to equivalent noise bandwidth of 0.8\,Hz).
The power spectral density of temporal current fluctuations is measured over a much larger bandwidth of 180\,kHz around 0.86\,MHz, using a homemade cryogenic amplifier and a tank circuit based on a superconducting coil.
The temperature of electrons in the devices is extracted from the quantum shot-noise across a quantum point contact (the right beam splitter QPC of the $L\simeq24\,\mu$m MZI set to $\tau_\mathrm{QPC}^\mathrm{R}\simeq0.5$).
See \cite{Iftikhar2016} for further details on the same experimental setup.

\subsection*{4. Central ohmic contacts characterization} 
The quality of the grounded central ohmic contact is characterized by the ratio of reflected over impinging current.
Ideally, there should be no reflected current.
In practice, if the impinging current is carried only by the outer edge channel (used for the interferometer), the reflected current is found to be negligible for both devices (below 1\%).
If the impinging current is carried by both the inner and outer edge channels, we find a reflected current in the range 11-21\% corresponding to a 22-42\% reflection of the inner edge channel from the central ohmic contact of the $L\simeq24\,\mu$m paths MZI, whereas for the $L\simeq0.1\,$mm MZI the reflected current remains essentially negligible (below 1\%).
Note that a good ohmic contact with the outer channel is assumed in Eq.~\ref{eqLphi} (an imperfect contact would further limit the amplitude of MZI oscillations).

\subsection*{5. Tunneling between inner and outer channels} 
Tunneling of electrons between adjacent, copropagating channels is usually negligible at filling factor $\nu=2$.
However, the propagation distances in the present devices can be considerable.
Following standard procedures \cite{vanWees1989a}, we determine the electron inter-channel tunneling along the MZI arms between the two QPC beam splitters, when the inner edge channel is not formed into small loops.
Note that the tunneling of electrons in the presence of small inner channel loops is expected to be much smaller because of the electronic level quantization within the loops and because of the Coulomb blockade of tunneling into (nearly) isolated islands (although this could not be measured because there is no dc current toward closed loops).
The tunneling between co-propagative inner and outer edge channels is obtained by applying a small bias selectively on one of the two channels, and by measuring at the end of the path the current in the other channel.
We find that the tunneling remains small for the $L\simeq24\,\mu$m MZI (between 2.5\% and 5\% [$\approx0\%$] of the injected current is detected on the second channel after propagating along the lower [upper] MZI arm). 
The tunneling is more important for the $L\simeq0.1\,$mm MZI (between 30\% and 48\% [between 10\% and 26\%] of the injected current is detected on the second channel after propagating along the lower [upper] MZI path).

\subsection*{6. Crosstalks characterization} 
Changing a gate voltage also slightly influences the other nearby gates. 
We take into account this small capacitive cross-talk correction on the beam splitter quantum point contacts (of at most 6\%, attained for the lateral plunger gate effect on the nearby left QPC).

\subsection*{7. Formation of inner channel loops} 
The comb shaped gates of homogeneous width (200\,nm) were polarized with a positive voltage of $+0.35\,$V during the cooldown from room temperature.
A broad gate voltage window is found to fully reflect the inner quantum Hall channel while completely letting through the outer channel (with a minimal common window from 0\,V to 0.13\,V, that applies simultaneously to each arm of both devices).
Such a behavior is usually observed on similar 2DEGs, thanks to the large energy separation between the two lowest Landau levels at filling factor $\nu=2$.
Note that the results corresponding to closed inner channel loops presented in the manuscript are not specific to a precise gate voltage setting (chosen within the minimal common window), but representative of the general behavior observed when the inner edge channel loops are completely closed while the outer edge channel is fully propagative.

\subsection*{8. Visibility of conductance oscillations versus QPC transmission} 
Here we provide more details on the procedure followed to extract the oscillations visibility data displayed in Fig.~3(a).
We performed relatively large plunger gate voltage sweeps, of 50\,mV corresponding to approximately 21 periods (with a step of 50\,$\mu$V corresponding to $1/46$ of a period), and repeated several times the same sweep (twice for the $L\simeq24\,\mu$m MZI, fourteen times for the more challenging $L\simeq0.1\,$mm MZI).
Each sweep was then decomposed into one-period intervals with half a period of overlap between consecutive intervals, and a `local' visibility of the oscillations in $\tau_\mathrm{MZI}$ was extracted from $\mathcal{V}\equiv(\tau_\mathrm{MZI}^\mathrm{max}-\tau_\mathrm{MZI}^\mathrm{min})/(\tau_\mathrm{MZI}^\mathrm{max}+\tau_\mathrm{MZI}^\mathrm{min})$ in each of these intervals.
The symbols in Fig.~3(a) display the many different values of $\mathcal{V}$ obtained by this procedure.

\subsection*{9. Temporal noise spectral density} 
Here we provide more details on the noise data and calculations displayed in Fig.~3(b).
The data points represent the excess power spectral density of the current detected on electrode D (see Fig.~1(a)), i.e. the total noise from which is subtracted the equilibrium noise offset at $V_\mathrm{dc}=0$ (that includes the contribution of the amplification chain).
To make sure that the noise dependence in the MZI quantum phase is fully averaged out, the displayed data represents the average of many noise measurements equally distributed in a range of plunger gate voltage corresponding to several periods (240 [40] values of $V_\mathrm{pl}$ distributed over approximately 5 [2] periods for the $L\simeq24~[100]\,\mu$m MZI).
The displayed calculations (continuous lines) only include the contribution of `slow' fluctuations in the MZI quantum phase $\delta\phi(t)$, detected within a 180\,kHz window around 0.86\,MHz, and not the quantum shot noise contribution further discussed below.
From the relationship $\tau_\mathrm{MZI}(t)=0.5\left(1+\mathcal{V}\sin\left[\langle\phi\rangle+\delta\phi(t)\right]\right)$, it is straightforward to obtain that the resulting noise in transmitted current is given by \cite{Marquardt2004}:
\begin{equation}
\langle I_\mathrm{\delta\phi}^2 \rangle_\propto\frac{V_\mathrm{dc}^2e^4}{h^2}\mathcal{V}^2, \label{eqPhaseNoise1}
\end{equation}
with $h$ the Planck constant and $e$ the elementary electron charge.
At low $V_\mathrm{dc}$ bias (as long as the oscillation visibility $\mathcal{V}$ is not significantly reduced), one thus expects a quadratic increase.
Using the relationship between visibility and phase coherence length given Eq.~\ref{eqLphi}, this expression becomes:
\begin{equation}
\langle I_\mathrm{\delta\phi}^2 \rangle\propto\frac{V_\mathrm{dc}^2e^4}{h^2}\exp\left(\frac{-4L}{L_\phi}\right). \label{eqPSDnoshotnoise}
\end{equation}
The calculations displayed as black continuous lines are obtained from Eq.~\ref{eqPSDnoshotnoise}, using for both devices the same empirical expression $L_\phi(V_\mathrm{dc})=0.25\,\mathrm{mm}\times\exp\left[-(V_\mathrm{dc}/26\,\mu\mathrm{V})^2\right]$ (displayed in the inset), the corresponding MZI length $L=24\,\mu$m or $0.1\,$mm, and where the unknown prefactor (depending on the number and coupling strength of the phase noise sources) is considered here as a free parameter for each device.
The smaller quantum shot noise contribution (not included in Eq.~\ref{eqPSDnoshotnoise}) is linear in $V_\mathrm{dc}$ and does not rely on the presence of MZI quantum interferences.
As expected if the vanishing current noise results from a quantum decoherence by `fast' phase fluctuations \cite{Marquardt2004} (compared to the electron quantum coherence), the amplitude of the linear noise is found strongly suppressed compared to the naive expectation $\langle I^2\rangle=2e(V_\mathrm{dc}e^2/h)\langle\tau_\mathrm{MZI}\rangle(1-\langle\tau_\mathrm{MZI}\rangle)$, by a factor of 4 (6) for the MZI of arm length $L\simeq24~(100)\,\mu$m.

\begin{figure}
\centering\includegraphics [width=86mm]{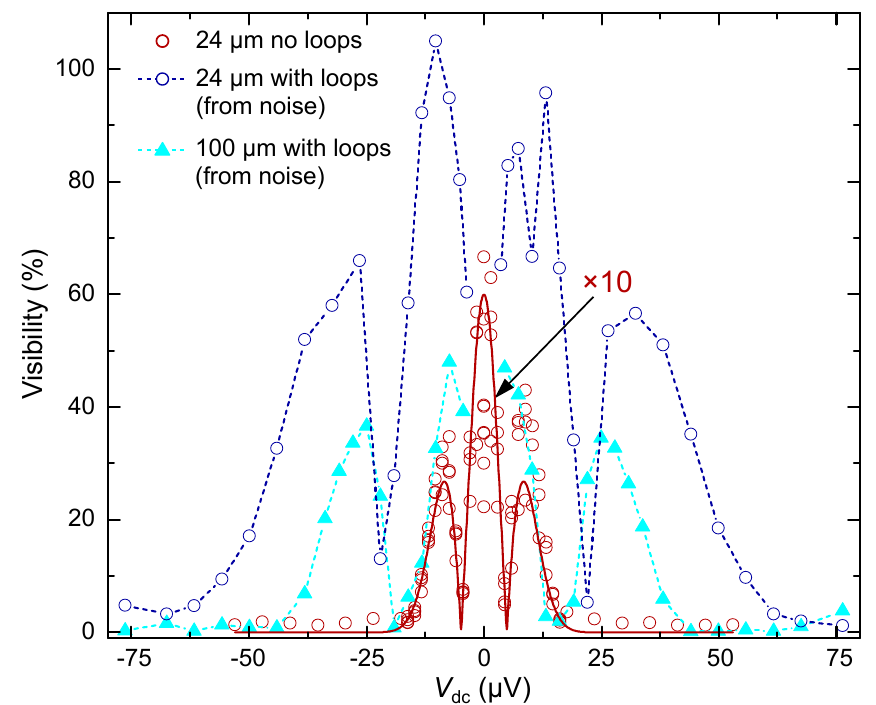}
\caption{
Out-of-equilibrium visibility in the differential current $\mathcal{V}_\mathrm{diff}$.
The red circles represent measurements of the visibility of the oscillations in the differential transmitted current across the $L\simeq24\,\mu$m MZI without inner channel loops, as a function of the applied dc bias voltage.
The continuous red line is calculated from Eq.~\ref{eqVisGaussian} (see text).
The dark blue circles (light blue full triangles) connected by dashed lines represent the differential visibility on the $L\simeq24$ $(100)\,\mu$m MZI with formed inner channel loops, which was extracted from the noise measurements displayed in Fig.~3(b) (see text).
}
\end{figure}

\subsection*{10. Comparison of voltage bias robustness with and without inner channel loops} 
In the absence of inner channel loops, the negligible MZI phase noise does not allow us to probe $L_\phi(V_\mathrm{dc})$ through the power spectral density of the transmitted current's temporal fluctuations.
However, on the $L\simeq24\,\mu$m MZI where quantum oscillations are visible without loops, it is possible to determine, versus dc voltage bias, their visibility $\mathcal{V}_\mathrm{diff}$ in the transmitted \textit{differential current} $dI_\mathrm{MZI}/dV_\mathrm{dc}$.
The `diff' subscript is introduced here to clearly distinguish between, on the one hand, this usually measured $\mathcal{V}_\mathrm{diff}$ and, on the other hand, the visibility $\mathcal{V}$ of oscillations in the total transmitted current $I_\mathrm{MZI}$ that is probed through noise measurements (Fig.~3(b)).
These two quantities are simply connected by the relation \cite{Roulleau2007}:
\begin{equation}
\mathcal{V}_\mathrm{diff}=|\mathcal{V}+V_\mathrm{dc}\partial\mathcal{V}/\partial V_\mathrm{dc}|.\label{eqVisACDC}
\end{equation}
Measurements of $\mathcal{V}_\mathrm{diff}(V_\mathrm{dc})$ on the $L\simeq24\,\mu$m MZI without loops are shown in Fig.~5 as open red circles.
We find that $\mathcal{V}_\mathrm{diff}$ displays a single side lobe, with a first minimum at $|V_\mathrm{dc}|\simeq5\,\mu$V, and becomes negligible, below our experimental resolution, at $|V_\mathrm{dc}|\gtrsim15\,\mu$V.
The data can be reproduced by the simple single side lobe expression derived in \cite{Roulleau2007} assuming a gaussian phase averaging (continuous line in Fig.~5):
\begin{equation}
\mathcal{V}_\mathrm{diff}^\mathrm{gaussian}=\mathcal{V}_0\left|1-\frac{V_\mathrm{dc}^2}{V_0^2}\right|\exp\left(-\frac{V_\mathrm{dc}^2}{2V_0^2}\right),\label{eqVisGaussian}
\end{equation} 
with $\mathcal{V}_0=0.06$ the zero bias visibility and $V_0=5\,\mu$V the characteristic voltage scale also corresponding to the position of the intermediate minimum.
In order to compare the robustness of MZI interferences with and without inner channel loops, we have converted the noise data in Fig.~3(b) into the corresponding $\mathcal{V}_\mathrm{diff}$.
The resulting $\mathcal{V}_\mathrm{diff}$ is displayed in Fig.~5 as open dark blue circles and full light blue triangles for, respectively, the $L\simeq24$ and $100\,\mu$m MZI with loops.
This conversion first involves the determination of $\mathcal{V}$ from Eq.~\ref{eqPhaseNoise1} (using the measured noise spectral density from which the linear shot noise contribution observed at large $V_\mathrm{dc}$ was subtracted).
The unknown proportionality coefficient in Eq.~\ref{eqPhaseNoise1} is fixed by adjusting the visibility at low bias with its direct $V_\mathrm{dc}\approx0$ measurement displayed in Fig.~3(a).
The resulting $\mathcal{V}$ is then injected into Eq.~\ref{eqVisACDC} to obtain $\mathcal{V}_\mathrm{diff}$.
Comparing the two data sets at the same $L\simeq24\,\mu$m (open circles), we find that the robustness of the MZI visibility with $V_\mathrm{dc}$ is approximately four times larger in the presence of loops (dark blue) than without them (red).


\begin{thebibliography}{55}
\expandafter\ifx\csname natexlab\endcsname\relax\def\natexlab#1{#1}\fi
\expandafter\ifx\csname url\endcsname\relax
  \def\url#1{\texttt{#1}}\fi
\expandafter\ifx\csname urlprefix\endcsname\relax\def\urlprefix{URL }\fi

\bibitem[{Nazarov \& Blanter(2009)}]{Nazarov2009}
Nazarov, Y. \& Blanter, Y.
\newblock \emph{Quantum Transport} (Cambridge University Press, 2009).

\bibitem[{van Wees \emph{et~al.}(1989{\natexlab{a}})}]{vanWees1989b}
van Wees, B.~J. \emph{et~al.}
\newblock Observation of zero-dimensional states in a one-dimensional electron
  interferometer.
\newblock \emph{Phys. Rev. Lett.} \textbf{62}, 2523--2526 (1989{\natexlab{a}}).

\bibitem[{{Ji} \emph{et~al.}(2003)}]{Ji2003}
{Ji}, Y. \emph{et~al.}
\newblock {An electronic Mach-Zehnder interferometer}.
\newblock \emph{Nature} \textbf{422}, 415--418 (2003).

\bibitem[{{Neder} \emph{et~al.}(2007)}]{Neder2007}
{Neder}, I. \emph{et~al.}
\newblock {Interference between two indistinguishable electrons from
  independent sources}.
\newblock \emph{Nature} \textbf{448}, 333--337 (2007).

\bibitem[{{Yamamoto} \emph{et~al.}(2012)}]{Yamamoto2012}
{Yamamoto}, M. \emph{et~al.}
\newblock {Electrical control of a solid-state flying qubit}.
\newblock \emph{Nature Nanotech.} \textbf{7}, 247--251 (2012).

\bibitem[{Bocquillon \emph{et~al.}(2013{\natexlab{a}})}]{Bocquillon2013a}
Bocquillon, E. \emph{et~al.}
\newblock Coherence and Indistinguishability of Single Electrons Emitted by
  Independent Sources.
\newblock \emph{Science} \textbf{339}, 1054--1057 (2013{\natexlab{a}}).

\bibitem[{Dubois \emph{et~al.}(2013)}]{Dubois2013}
Dubois, J. \emph{et~al.}
\newblock Minimal-excitation states for electron quantum optics using levitons.
\newblock \emph{Nature} \textbf{502}, 659--663 (2013).

\bibitem[{Bertoni \emph{et~al.}(2000)Bertoni, Bordone, Brunetti, Jacoboni \&
  Reggiani}]{Bertoni2000}
Bertoni, A., Bordone, P., Brunetti, R., Jacoboni, C. \& Reggiani, S.
\newblock Quantum Logic Gates based on Coherent Electron Transport in Quantum
  Wires.
\newblock \emph{Phys. Rev. Lett.} \textbf{84}, 5912--5915 (2000).

\bibitem[{{Ionicioiu} \emph{et~al.}(2001){Ionicioiu}, {Amaratunga} \&
  {Udrea}}]{Ionicioiu2001}
{Ionicioiu}, R., {Amaratunga}, G. \& {Udrea}, F.
\newblock {Quantum Computation with Ballistic Electrons}.
\newblock \emph{Int. J Mod. Phys. B} \textbf{15}, 125--133 (2001).

\bibitem[{Stace \emph{et~al.}(2004)Stace, Barnes \& Milburn}]{Stace2004}
Stace, T., Barnes, C. \& Milburn, G.
\newblock Mesoscopic One-Way Channels for Quantum State Transfer via the
  Quantum Hall Effect.
\newblock \emph{Phys. Rev. Lett.} \textbf{93}, 126804 (2004).

\bibitem[{Glattli \& Roulleau(2017)}]{Glattli2017}
Glattli, D. \& Roulleau, P.
\newblock Levitons for electron quantum optics.
\newblock \emph{Phys. Status Solidi B} \textbf{254}, 1600650 (2017).

\bibitem[{{B{\"a}uerle} \emph{et~al.}(2018)}]{Bauerle2018}
{B{\"a}uerle}, C. \emph{et~al.}
\newblock {Coherent control of single electrons: a review of current progress}.
\newblock \emph{Rep. Prog. Phys.} \textbf{81}, 056503 (2018).

\bibitem[{Pierre \emph{et~al.}(2003)}]{Pierre2003}
Pierre, F. \emph{et~al.}
\newblock Dephasing of electrons in mesoscopic metal wires.
\newblock \emph{Phys. Rev. B} \textbf{68}, 085413 (2003).

\bibitem[{Niimi \emph{et~al.}(2010)}]{Niimi2010}
Niimi, Y. \emph{et~al.}
\newblock Quantum coherence at low temperatures in mesoscopic systems: Effect
  of disorder.
\newblock \emph{Phys. Rev. B} \textbf{81}, 245306 (2010).

\bibitem[{Miao \emph{et~al.}(2007)}]{Miao2007}
Miao, F. \emph{et~al.}
\newblock Phase-Coherent Transport in Graphene Quantum Billiards.
\newblock \emph{Science} \textbf{317}, 1530--1533 (2007).

\bibitem[{Roulleau \emph{et~al.}(2008{\natexlab{a}})}]{Roulleau2008a}
Roulleau, P. \emph{et~al.}
\newblock Direct Measurement of the Coherence Length of Edge States in the
  Integer Quantum Hall Regime.
\newblock \emph{Phys. Rev. Lett.} \textbf{100}, 126802 (2008{\natexlab{a}}).

\bibitem[{Litvin \emph{et~al.}(2008)Litvin, Helzel, Tranitz, Wegscheider \&
  Strunk}]{Litvin2008}
Litvin, L.~V., Helzel, A., Tranitz, H.-P., Wegscheider, W. \& Strunk, C.
\newblock Edge-channel interference controlled by Landau level filling.
\newblock \emph{Phys. Rev. B} \textbf{78}, 075303 (2008).

\bibitem[{Gurman \emph{et~al.}(2016)Gurman, Sabo, Heiblum, Umansky \&
  Mahalu}]{Gurman2016}
Gurman, I., Sabo, R., Heiblum, M., Umansky, V. \& Mahalu, D.
\newblock Dephasing of an electronic two-path interferometer.
\newblock \emph{Phys. Rev. B} \textbf{93}, 121412 (2016).

\bibitem[{Levkivskyi \& Sukhorukov(2008)}]{Levkivskyi2008}
Levkivskyi, I.~P. \& Sukhorukov, E.~V.
\newblock Dephasing in the electronic Mach-Zehnder interferometer at filling
  factor $\ensuremath{\nu}=2$.
\newblock \emph{Phys. Rev. B} \textbf{78}, 045322 (2008).

\bibitem[{Roulleau \emph{et~al.}(2008{\natexlab{b}})}]{Roulleau2008b}
Roulleau, P. \emph{et~al.}
\newblock Noise Dephasing in Edge States of the Integer Quantum Hall Regime.
\newblock \emph{Phys. Rev. Lett.} \textbf{101}, 186803 (2008{\natexlab{b}}).

\bibitem[{Altimiras \emph{et~al.}(2010)}]{Altimiras2010}
Altimiras, C. \emph{et~al.}
\newblock Tuning Energy Relaxation along Quantum Hall Channels.
\newblock \emph{Phys. Rev. Lett.} \textbf{105}, 226804 (2010).

\bibitem[{{Cabart} \emph{et~al.}(2018){Cabart}, {Roussel}, {F{\`e}ve} \&
  {Degiovanni}}]{Cabart2018}
{Cabart}, C., {Roussel}, B., {F{\`e}ve}, G. \& {Degiovanni}, P.
\newblock {Taming electronic decoherence in 1D chiral ballistic quantum
  conductors}.
\newblock \emph{ArXiv} 1804.04054 (2018).

\bibitem[{Bocquillon \emph{et~al.}(2014)}]{Bocquillon2014}
Bocquillon, E. \emph{et~al.}
\newblock Electron quantum optics in ballistic chiral conductors.
\newblock \emph{Ann. Phys. (Berlin)} \textbf{526}, 1--30 (2014).

\bibitem[{Bautze \emph{et~al.}(2014)}]{Bautze2014}
Bautze, T. \emph{et~al.}
\newblock Theoretical, numerical, and experimental study of a flying qubit
  electronic interferometer.
\newblock \emph{Phys. Rev. B} \textbf{89}, 125432 (2014).

\bibitem[{{Duprez} \emph{et~al.}(2019)}]{Duprez2019b}
{Duprez}, H. \emph{et~al.}
\newblock {Transferring the quantum state of electrons across a Fermi sea with
  Coulomb interaction}.
\newblock \emph{arXiv e-prints} arXiv:1902.07569 (2019).

\bibitem[{Wei \emph{et~al.}(2017)}]{Wei2017}
Wei, D. \emph{et~al.}
\newblock Mach-Zehnder interferometry using spin- and valley-polarized quantum
  Hall edge states in graphene.
\newblock \emph{Sci. Adv.} \textbf{3}, e1700600 (2017).

\bibitem[{Giamarchi(2003)}]{Giamarchi2003}
Giamarchi, T.
\newblock \emph{Quantum Physics in One Dimension} (Oxford University Press,
  2003).

\bibitem[{le~Sueur \emph{et~al.}(2010)}]{leSueur2010}
le~Sueur, H. \emph{et~al.}
\newblock Energy Relaxation in the Integer Quantum Hall Regime.
\newblock \emph{Phys. Rev. Lett.} \textbf{105}, 056803 (2010).

\bibitem[{Itoh \emph{et~al.}(2018)}]{Itoh2018}
Itoh, K. \emph{et~al.}
\newblock Signatures of a Nonthermal Metastable State in Copropagating Quantum
  Hall Edge Channels.
\newblock \emph{Phys. Rev. Lett.} \textbf{120}, 197701 (2018).

\bibitem[{Bocquillon \emph{et~al.}(2013{\natexlab{b}})}]{Bocquillon2013b}
Bocquillon, E. \emph{et~al.}
\newblock Separation of neutral and charge modes in one-dimensional chiral edge
  channels.
\newblock \emph{Nat. Commun.} \textbf{4}, 1839 (2013{\natexlab{b}}).

\bibitem[{Inoue \emph{et~al.}(2014{\natexlab{a}})}]{Inoue2014b}
Inoue, H. \emph{et~al.}
\newblock Charge Fractionalization in the Integer Quantum Hall Effect.
\newblock \emph{Phys. Rev. Lett.} \textbf{112}, 166801 (2014{\natexlab{a}}).

\bibitem[{Hashisaka \emph{et~al.}(2017)Hashisaka, Hiyama, Akiho, Muraki \&
  Fujisawa}]{Hashisaka2017}
Hashisaka, M., Hiyama, N., Akiho, T., Muraki, K. \& Fujisawa, T.
\newblock Waveform measurement of charge- and spin-density wavepackets in a
  chiral Tomonaga-Luttinger liquid.
\newblock \emph{Nat. Phys.} \textbf{13}, 559--562 (2017).

\bibitem[{Marguerite \emph{et~al.}(2016)}]{Marguerite2016}
Marguerite, A. \emph{et~al.}
\newblock Decoherence and relaxation of a single electron in a one-dimensional
  conductor.
\newblock \emph{Phys. Rev. B} \textbf{94}, 115311 (2016).

\bibitem[{Huynh \emph{et~al.}(2012)}]{Huynh2012}
Huynh, P.-A. \emph{et~al.}
\newblock Quantum Coherence Engineering in the Integer Quantum Hall Regime.
\newblock \emph{Phys. Rev. Lett.} \textbf{108}, 256802 (2012).

\bibitem[{Marquardt \& Bruder(2004)}]{Marquardt2004}
Marquardt, F. \& Bruder, C.
\newblock Effects of dephasing on shot noise in an electronic Mach-Zehnder
  interferometer.
\newblock \emph{Phys. Rev. B} \textbf{70}, 125305 (2004).

\bibitem[{{Kr{\"a}henmann} \emph{et~al.}(2019)}]{Krahenman2019}
{Kr{\"a}henmann}, T. \emph{et~al.}
\newblock {Auger-spectroscopy in quantum Hall edge channels: a possible
  resolution to the missing energy problem}.
\newblock \emph{arXiv e-prints} arXiv:1902.10065 (2019).

\bibitem[{Sukhorukov \& Cheianov(2007)}]{Sukhorukov2007}
Sukhorukov, E. \& Cheianov, V.
\newblock Resonant Dephasing in the Electronic Mach-Zehnder Interferometer.
\newblock \emph{Phys. Rev. Lett.} \textbf{99}, 156801 (2007).

\bibitem[{Chalker \emph{et~al.}(2007)Chalker, Gefen \& Veillette}]{Chalker2007}
Chalker, J., Gefen, Y. \& Veillette, M.
\newblock Decoherence and interactions in an electronic Mach-Zehnder
  interferometer.
\newblock \emph{Phys. Rev. B} \textbf{76}, 085320 (2007).

\bibitem[{Kovrizhin \& Chalker(2010)}]{Kovrizhin2010}
Kovrizhin, D. \& Chalker, J.
\newblock Multiparticle interference in electronic Mach-Zehnder
  interferometers.
\newblock \emph{Phys. Rev. B} \textbf{81}, 155318 (2010).

\bibitem[{Schneider \emph{et~al.}(2011)Schneider, Bagrets \&
  Mirlin}]{Schneider2011}
Schneider, M., Bagrets, D. \& Mirlin, A.
\newblock Theory of the nonequilibrium electronic Mach-Zehnder interferometer.
\newblock \emph{Phys. Rev. B} \textbf{84}, 075401 (2011).

\bibitem[{Hashisaka \emph{et~al.}(2012)Hashisaka, Washio, Kamata, Muraki \&
  Fujisawa}]{Hashisaka2012}
Hashisaka, M., Washio, K., Kamata, H., Muraki, K. \& Fujisawa, T.
\newblock Distributed electrochemical capacitance evidenced in high-frequency
  admittance measurements on a quantum Hall device.
\newblock \emph{Phys. Rev. B} \textbf{85}, 155424 (2012).

\bibitem[{Tu \emph{et~al.}(2018)}]{Tu2018}
Tu, N. \emph{et~al.}
\newblock Coupling between quantum Hall edge channels on opposite sides of a
  Hall bar.
\newblock \emph{Solid State Commun.} \textbf{283}, 32--36 (2018).

\bibitem[{MacDonald \emph{et~al.}(1993)MacDonald, Yang \&
  Johnson}]{MacDonald1993}
MacDonald, A., Yang, S. \& Johnson, M.
\newblock {Quantum dots in strong magnetic fields: Stability criteria for the
  maximum density droplet}.
\newblock \emph{Aust. J. Phys.} \textbf{46}, 345 (1993).

\bibitem[{Chamon \& Wen(1994)}]{Chamon1994}
Chamon, C. \& Wen, X.
\newblock Sharp and smooth boundaries of quantum Hall liquids.
\newblock \emph{Phys. Rev. B} \textbf{49}, 8227--8241 (1994).

\bibitem[{Aleiner \& Glazman(1994)}]{Aleiner1994}
Aleiner, I. \& Glazman, L.
\newblock Novel edge excitations of two-dimensional electron liquid in a
  magnetic field.
\newblock \emph{Phys. Rev. Lett.} \textbf{72}, 2935--2938 (1994).

\bibitem[{Jezouin \emph{et~al.}(2013)}]{Jezouin2013b}
Jezouin, S. \emph{et~al.}
\newblock Quantum Limit of Heat Flow Across a Single Electronic Channel.
\newblock \emph{Science} \textbf{342}, 601--604 (2013).

\bibitem[{{Banerjee} \emph{et~al.}(2017)}]{Banerjee2017}
{Banerjee}, M. \emph{et~al.}
\newblock Observed Quantization of Anyonic Heat Flow.
\newblock \emph{Nature} \textbf{545}, 75--79 (2017).

\bibitem[{Sivre \emph{et~al.}(2018)}]{Sivre2018}
Sivre, E. \emph{et~al.}
\newblock Heat Coulomb blockade of one ballistic channel.
\newblock \emph{Nat. Phys.} \textbf{14}, 145--148 (2018).

\bibitem[{Altimiras \emph{et~al.}(2012)}]{Altimiras2012}
Altimiras, C. \emph{et~al.}
\newblock Chargeless Heat Transport in the Fractional Quantum Hall Regime.
\newblock \emph{Phys. Rev. Lett.} \textbf{109}, 026803 (2012).

\bibitem[{{Venkatachalam} \emph{et~al.}(2012){Venkatachalam}, {Hart},
  {Pfeiffer}, {West} \& {Yacoby}}]{Venkatachalam2012}
{Venkatachalam}, V., {Hart}, S., {Pfeiffer}, L., {West}, K. \& {Yacoby}, A.
\newblock {Local thermometry of neutral modes on the quantum Hall edge}.
\newblock \emph{Nat. Phys.} \textbf{8}, 676--681 (2012).

\bibitem[{Inoue \emph{et~al.}(2014{\natexlab{b}})}]{Inoue2014a}
Inoue, H. \emph{et~al.}
\newblock Proliferation of neutral modes in fractional quantum Hall states.
\newblock \emph{Nat. Commun.} \textbf{5}, 4067 (2014{\natexlab{b}}).

\bibitem[{Chklovskii \emph{et~al.}(1992)Chklovskii, Shklovskii \&
  Glazman}]{Chklovskii1992}
Chklovskii, D.~B., Shklovskii, B.~I. \& Glazman, L.~I.
\newblock Electrostatics of edge channels.
\newblock \emph{Phys. Rev. B} \textbf{46}, 4026--4034 (1992).

\bibitem[{Iftikhar \emph{et~al.}(2016)}]{Iftikhar2016}
Iftikhar, Z. \emph{et~al.}
\newblock Primary thermometry triad at 6 mK in mesoscopic circuits.
\newblock \emph{Nat. Commun.} \textbf{7}, 12908 (2016).

\bibitem[{van Wees \emph{et~al.}(1989{\natexlab{b}})}]{vanWees1989a}
van Wees, B. \emph{et~al.}
\newblock {Anomalous integer quantum Hall effect in the ballistic regime with
  quantum point contacts}.
\newblock \emph{Phys. Rev. Lett.} \textbf{62}, 1181--1184 (1989{\natexlab{b}}).

\bibitem[{Roulleau \emph{et~al.}(2007)}]{Roulleau2007}
Roulleau, P. \emph{et~al.}
\newblock Finite bias visibility of the electronic Mach-Zehnder interferometer.
\newblock \emph{Phys. Rev. B} \textbf{76}, 161309 (2007).

\end{thebibliography}
\end{document}